\documentclass[11pt]{article}
\usepackage{pos}
\usepackage[utf8]{inputenc}
\usepackage{braket}
\usepackage{ascmac}
\usepackage{color}
\usepackage{here}
\usepackage{slashed}
\usepackage{url}
\usepackage{mathrsfs}
\usepackage{graphicx}
\usepackage[symbol]{footmisc}

\newcommand{\nn}{\nonumber}
\newcommand{\del}{\partial}

\numberwithin{equation}{section}

\title{Grassmann Tensor Renormalization Group for two-flavor massive Schwinger model with a theta term}
\ShortTitle{Grassmann TRG for $N_f=2$ massive Schwinger model with a $\theta$ term}

\author*[a]{Hayato Kanno}
\author[b,c]{Shinichiro Akiyama}
\author[d,e]{Kotaro Murakami}
\author[f]{Shinji Takeda}

\affiliation[a]{RIKEN BNL Research Center,\\
Brookhaven National Laboratory, Upton, NY 11973, USA}

\affiliation[b]{Center for Computational Sciences, University of Tsukuba,\\
1-1-1 Tennodai, Tsukuba, Ibaraki 305-8577, Japan}

\affiliation[c]{Graduate School of Science, The University of Tokyo,\\
7-3-1 Hongo, Bunkyo-ku, Tokyo 113-0033, Japan}

\affiliation[d]{Department of Physics, Institute of Science Tokyo,\\
2-12-1 Ookayama, Megro, Tokyo 152-8551, Japan}

\affiliation[e]{Interdisciplinary Theoretical and Mathematical Sciences Program (iTHEMS), RIKEN,\\
2-1 Hirosawa, Wako, Saitama 351-0198, Japan}

\affiliation[f]{Institute for Theoretical Physics, Kanazawa University,\\
Kakuma-machi, Kanazawa 920-1192, Japan}

\emailAdd{hayato.kanno@riken.jp}
\emailAdd{akiyama@ccs.tsukuba.ac.jp}
\emailAdd{kotaro.murakami@yukawa.kyoto-u.ac.jp}
\emailAdd{takeda@hep.s.kanazawa-u.ac.jp}

\abstract{We investigate the $N_f=2$ Schwinger model with the massive staggered fermions in the presence of a $2\pi$ periodic $\theta$ term, using the Grassmann tensor renormalization group.
Thanks to the Grassmann tensor network formulation, there is no difficulty in dealing with the massive staggered fermions. 
We study the $\theta$ dependence of the free energy in the thermodynamic limit. 
Our calculation provides consistent results with the analytical solution in the large mass limit. 
The results also suggest that the $N_f=2$ Schwinger model on a lattice has a different phase structure from that described by the continuum theory.
}

\notes{
\note{UTHEP-798, UTCCS-P-162, RIKEN-iTHEMS-Report-25}
}

\FullConference{The 41st International Symposium on Lattice Field Theory (LATTICE2024)\\
 28 July - 3 August 2024\\
Liverpool, UK\\}

\begin{document}
\maketitle

\section{Introduction}
Tensor network methods have become prominent candidates for avoiding the infamous sign problem in the Monte Carlo (MC) simulation.
Along with the famous method, the density matrix renormalization group (DMRG)~\cite{White:1992zz}, Lagrangian-based approaches such as the tensor renormalization group~\cite{Levin:2006jai}, which is a variant of the real-space renormalization group employing the tensor network representation of path integrals, have been developed and shown to be efficient for various quantum statistical models and field theories in recent years.

In this paper, we numerically study the Schwinger model~\cite{Schwinger:1962tp}, the two-dimensional (2D) Quantum ElectroDynamics (QED), with a $\theta$ term. 
The model is well known as a toy model of the 4D Quantum ChromoDynamics (QCD) and has a non-trivial topological nature related to a $\theta$ term.
In particular, to understand the vacuum structure in situations more similar to QCD, we consider the two-flavor ($N_f=2$) Schwinger model.
The nature of vacuum at $\theta=\pi$ is quite different between the cases with $N_{f}=1$ and $N_{f}\ge2$; the twofold vacuum appears for a fermion mass larger than a certain value with $N_{f}=1$ while the vacuum is expected to emerge for any positive mass with $N_f\ge2$.
The model with $N_{f}=2$ is studied analytically in the heavy and small mass limits, where for the latter case the mass perturbation~\cite{Coleman:1975pw, Coleman:1976uz} works well.
However, the model is not generally solvable when the mass is finite. 
Therefore, the numerical study is of essential importance for this model.
Several tensor network studies on this model have recently been reported ~\cite{Dempsey:2023gib, Itou:2024psm}.

The TRG method has various advantages over conventional approaches.
First, we can take a lattice volume large enough to be identified as the thermodynamic limit for reasonable computational cost.
Second, the TRG enables us to work on a torus straightforwardly, in which a $\theta$ parameter is $2\pi$ periodic.
In addition, we can directly treat the Grassmann variables~\cite{Gu:2010yh,Shimizu:2014uva,Akiyama:2020sfo}, that is, the fermion fields, in the tensor network representation of path integrals. 
Taking these advantages, we utilize the TRG to calculate the free energy density of the massive $N_{f}=2$ Schwinger model with a $2\pi$ periodic $\theta$ term. 
This article is based on the paper \cite{Kanno:2024elz}.



\section{Schwinger model in continuum}
\label{sec:review}

The continuum action of the Schwinger model is given by
\begin{align}
    S =& \int {\rm d}^2x \bigg\{\frac{1}{4g^2}F_{\mu\nu}F^{\mu\nu} + \frac{{\rm i}\theta}{4\pi}\epsilon^{\mu\nu}F_{\mu\nu} + \bar{\psi}{\rm i}\gamma^{\mu}(\del_{\mu}+{\rm i}A_{\mu})\psi + m\bar{\psi}\psi  \bigg\} \ ,
\label{contS}
\end{align}
where $A_{\mu}$ is $U(1)$ gauge field, and $\psi$ and $\bar{\psi}$ are Dirac fermions in fundamental representation.
Here, $m$($\ge0$) is the fermion mass parameter and $g$ is the gauge coupling.
The second term in the above equation corresponds to the $\theta$ term in this model.
Hereafter, we mainly focus on the $\theta$ dependence of the free energy densities for $N_f=2$.
The free energy density with a dimensionless combination in the large mass limit can be calculated analytically as
\begin{align}
    -\frac{\log Z(\theta)}{g^2 V} =&  \min_n  \frac{1}{8\pi^2}\left(\theta -2\pi n \right)^2 \ .
    \label{eq:f_Maxwell_result}
\end{align}
On the other hand, the energy density in the small mass region is obtained via the mass perturbation~\cite{Coleman:1976uz} as
\begin{align}
    -\frac{\log Z(\theta)}{g^2 V}
    =& \min_n \left\{\left(\mathrm{e}^{\gamma}\right)^{\frac{4}{3}}\pi^{-\frac{5}{3}}2^{\frac{1}{3}}\left(\frac{m^2}{g^2}\right)^{\frac{2}{3}}\cos^{\frac{4}{3}}\left(\frac{\theta-2\pi n}{2}\right) \right\} \ ,
    \label{eq:f_exact_light_result}
\end{align}
with the Euler constant $\gamma = 0.577\cdots$. 
These two expressions are evaluated in the thermodynamic limit.
For the detailed derivations, see Ref.~\cite{Kanno:2024elz}. 


\section{Lattice Schwinger model and its tensor network representation}
\label{sec:lattice}


To simulate the lattice model corresponding to Eq.~\eqref{contS}, we employ the usual $U(1)$ Wilson gauge action with the $2\pi$-periodic $\theta$ term and staggered fermion action: 
\begin{align}
\label{eq:action}
	S=
	&-\beta\sum_{n\in\Lambda_{2}}
	\Re\left[U_{1}(n)U_{2}(n+\hat{1})U^{*}_{1}(n+\hat{2})U^{*}_{2}(n)\right]
    -\frac{\theta}{2\pi}\sum_{n}
	\log\left[U_{1}(n)U_{2}(n+\hat{1})U^{*}_{1}(n+\hat{2})U^{*}_{2}(n)\right] \nn\\
	&+\frac{1}{2}\sum_{n}\sum_{\mu=1}^{2}
	\eta_{\mu}(n)\left[\bar{\chi}(n)U_{\mu}(n)\chi(n+\hat{\mu})-\bar{\chi}(n+\hat{\mu})U^{*}_{\mu}(n)\chi(n)\right]
	+m_0\sum_{n}\bar{\chi}(n)\chi(n)
	.
\end{align}
Here, $U_{\mu}(n)\in U(1)$ is the link variable, and $\chi(n)$ and $\bar{\chi}(n)$ are the single-component Grassmann variables. 
The mass parameter and the inverse gauge coupling are denoted by $m_0=am$ and $\beta=1/(ag)^2$, respectively, with a lattice spacing $a$.
We define the staggered sign function as $\eta_{1}(n)=1$ and $\eta_{2}(n)=(-1)^{n_{1}}$.
We assume the periodic (anti-periodic) boundary condition for the staggered fermions in $\hat{1}$ ($\hat{2}$) direction.
Parametrizing the link variable by the compact boson as $U_{\mu}(n)={\rm e}^{{\rm i}A_{\mu}(n)}$, the partition function is represented as
\begin{align}
\label{eq:lat_path_int}
	Z=
	\left(\prod_{n}\int\int{\rm d}\bar{\chi}(n){\rm d}\chi(n)\right)
	\left(\prod_{n,\mu}\int^{\pi}_{-\pi}\frac{{\rm d}A_{\mu}(n)}{2\pi}\right)
	{\rm e}^{-S}
	.
\end{align}


The tensor network representation of Eq.~\eqref{eq:lat_path_int} is derived as follows.
First, we use the Gauss-Legendre quadrature following Ref.~\cite{Kuramashi:2019cgs} to replace each integral over $A_{\mu}(n)$ with the summation over finite numbers of sampling points according to
\begin{align}
\label{eq:def_gl_quadrature}
	\int^{1}_{-1}
        {\rm d}x_{\mu}(n)
        f\left(x_{\mu}(n)\right)
	\simeq
	\sum_{a_{\mu}(n)\in D_{K}}
        w_{a_{\mu}(n)}
        f\left(a_{\mu}(n)\right),
\end{align}
where $x_{\mu}(n)=A_{\mu}(n)/\pi$ and $f$ represents the corresponding integrand.
A set of $K$ sampling points is denoted by $D_{K}$.
Both $a_{\mu}(n)$ and $w_{a_{\mu}(n)}$ are determined by the Gauss-Legendre quadrature.
Using Eq.~\eqref{eq:def_gl_quadrature}, we can straightforwardly define a four-leg tensor on each plaquette as
\begin{align}
\label{eq:gauge_tensor}
	&T^{(g)}_{a_{2}(n+\hat{1})a_{1}(n+\hat{2})a_{2}(n)a_{1}(n)}
	=
	\frac{\sqrt{w_{a_{1}(n)}w_{a_{2}(n+\hat{1})}w_{a_{1}(n+\hat{2})}w_{a_{2}(n)}}}{4}
	\nonumber\\
	&\times
	{\rm e}^{
            \beta\cos\left[
                \pi\left(a_{1}(n)+a_{2}(n+\hat{1})-a_{1}(n+\hat{2})-a_{2}(n)\right)
            \right]
            +
            \frac{\theta}{2\pi}\log\left[
                {\rm e}^{{\rm i}\pi\left(a_{1}(n)+a_{2}(n+\hat{1})-a_{1}(n+\hat{2})-a_{2}(n)\right)}
            \right]
        }
	,
\end{align}
whose bond dimension is $K$.
Next, we employ the Grassmann tensor network formulation~\cite{Akiyama:2020sfo} to deal with the staggered fermions and derive the fundamental Grassmann tensor $T^{(f)}_n$ at each lattice site $n$. 
See Ref.~\cite{Kanno:2024elz} for the details of $T^{(f)}_n$.
Introducing the auxiliary fermion fields, $T^{(f)}_n$ is defined as a four-leg tensor with bond dimension 4.
Combining $T^{(g)}_n$ in Eq.~\eqref{eq:gauge_tensor} and the Grassmann tensor $T^{(f)}_n$, the path integral in Eq.~\eqref{eq:lat_path_int} is now expressed by
\begin{align}
\label{eq:gtn}
    Z\simeq Z(K)={\rm gTr}\left[\prod_{n}T^{(g)}_nT^{(f)}_n\right].
\end{align}
The ``gTr" denotes the integration over the auxiliary Grassmann fields and the summation over the discretized gauge fields~\cite{Akiyama:2024ush}.
The right-hand side of the above equation represents the Grassmann tensor network with the bond dimension $4K$.
To evaluate Eq.~\eqref{eq:gtn}, we employ the bond-weighted TRG (BTRG) algorithm~\cite{PhysRevB.105.L060402,Akiyama:2022pse}, which improves the accuracy of the Levin-Nave TRG~\cite{Levin:2006jai} without increasing the computational cost.

We note that the Schwinger model with the massless staggered fermions was previously studied by the TRG approach in Ref.~\cite{Butt:2019uul}, where the tensor network representation was derived based on the world-line formulation~\cite{Gattringer:2015nea,Goschl:2017kml}. This formulation, however, introduces a non-local sign factor in the massive case, which makes it harder to perform the coarse-graining transformation in the TRG method.
In contrast, our tensor network is constructed as the Grassmann tensor network and there is no such difficulty even when the fermion mass is finite.


\section{Numerical results}
\label{sec:results}

We set $K\le25$ and $D\le120$, which are large enough to investigate the parameter region we will focus on.
For these algorithmic parameter dependencies, see Ref.~\cite{Kanno:2024elz}.
We define the dimensionless free energy density by
\begin{align}
    \label{eq:dimeless_f}
    f
    =-\frac{1}{g^2V}\log Z
    =-\frac{\beta}{L^2}\log Z , 
\end{align}    
where $L$ is the linear system size in the lattice unit ($V=a^2L^2$).



\begin{figure}[htbp]
    \centering
    \includegraphics[width=0.7\hsize]{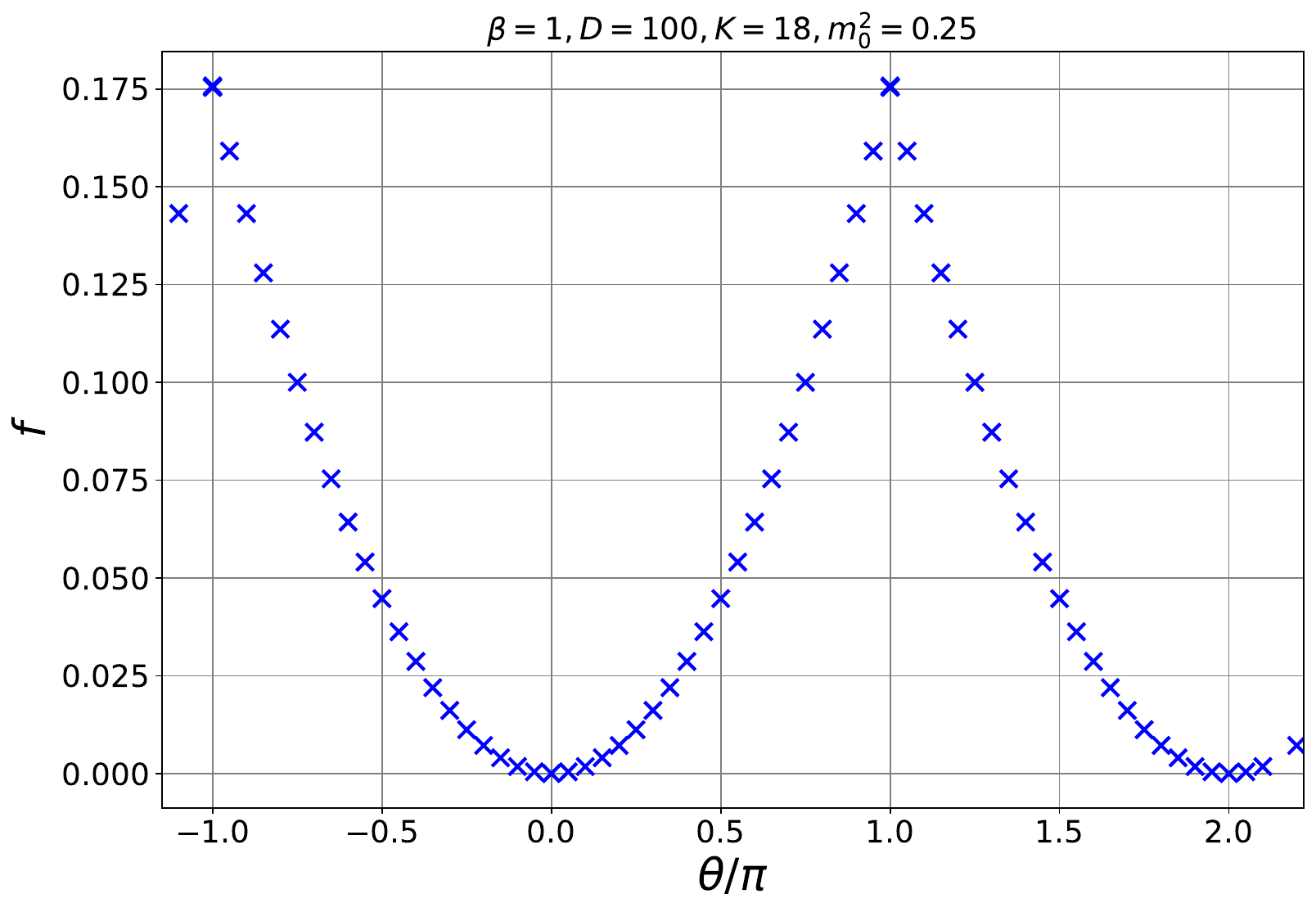}
    \caption{
        Free energy density as a function of $\theta/\pi$.
    }
    \label{fig:wide}
\end{figure}

Figure ~\ref{fig:wide} shows the thermodynamic free energy density at $\beta=1$, $m_{0}^{2}=0.25$ as a function of $\theta$. 
We have normalized the partition function by the value at $\theta=0$. 
The resulting free energy density shows a clear $2\pi$ periodicity associated with $\theta$. 
Hereafter, we show the results in the range of $\theta\in[0,\pi]$.

\begin{figure}[htbp]
    \centering
    \includegraphics[width=0.48\hsize]{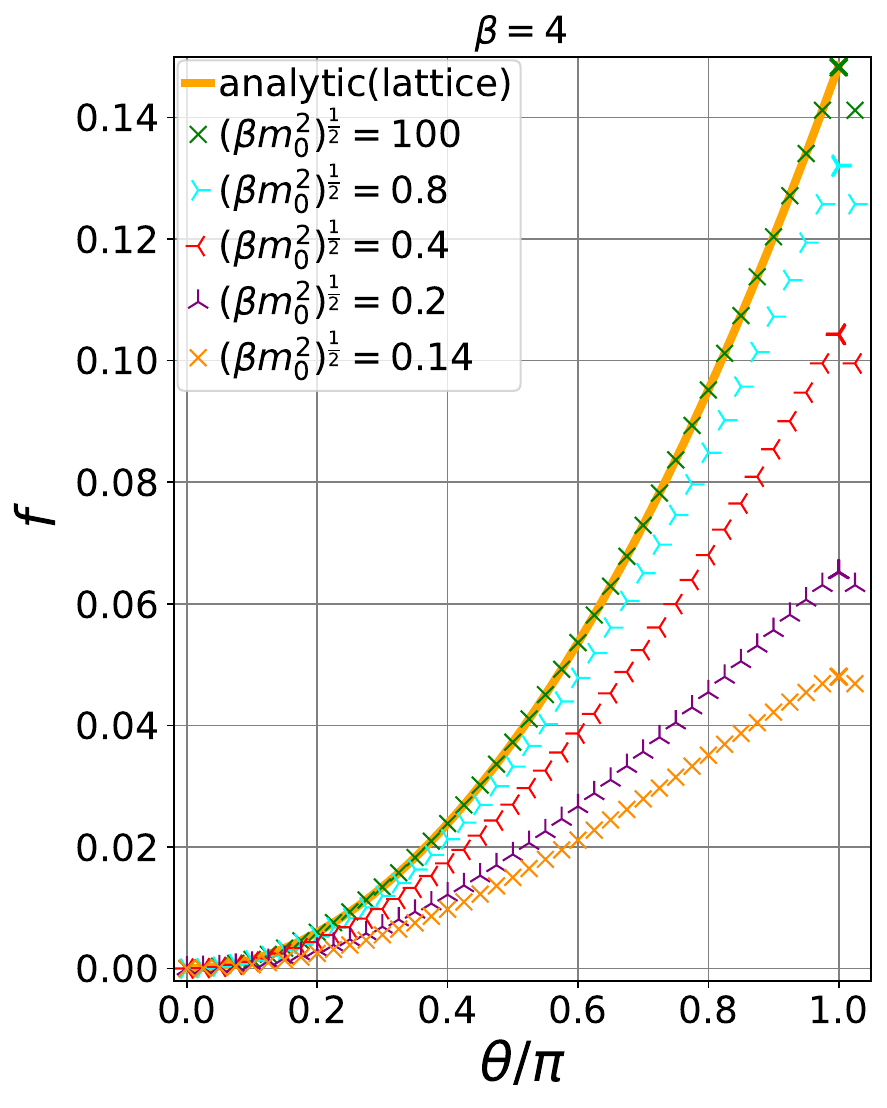}
    \includegraphics[width=0.5\hsize]{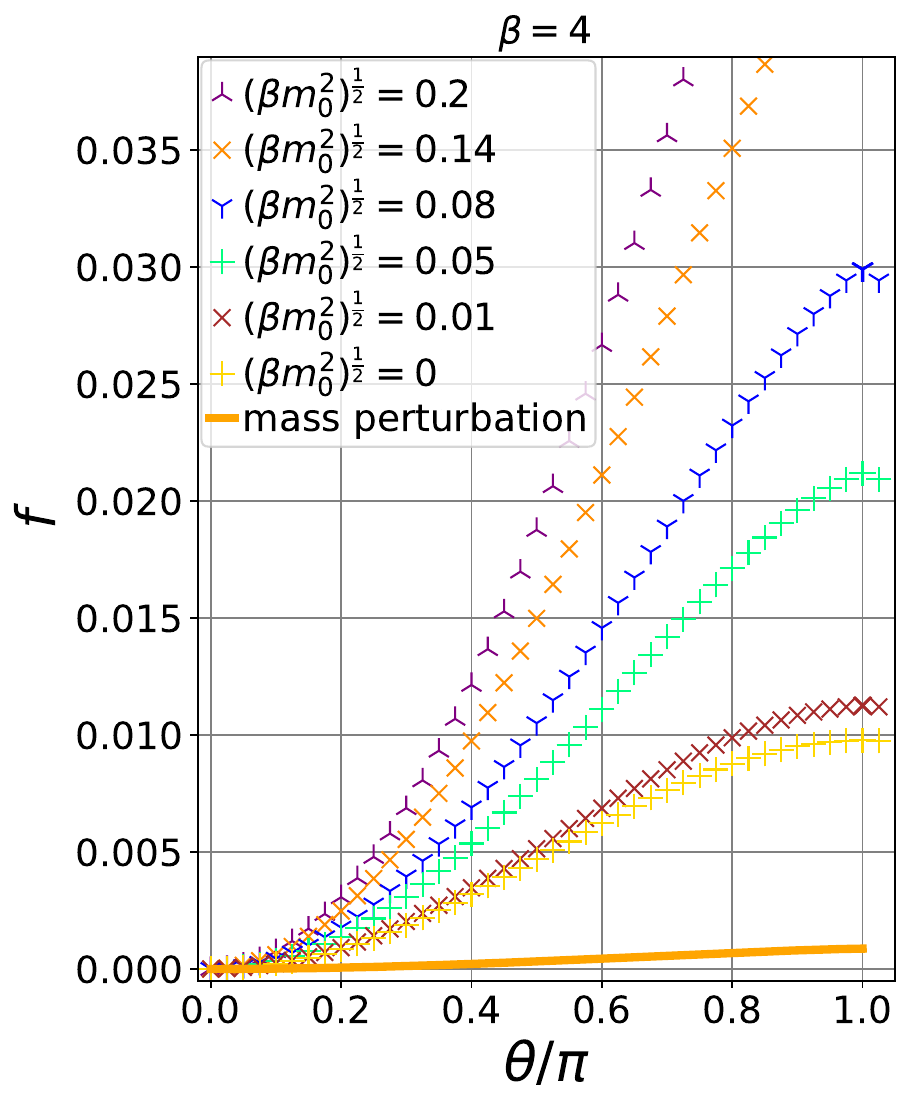}
    \caption{
        Free energy density as a function of $\theta/\pi$ with $\sqrt{\beta m_0^2} \geq 0.14$ (left) and $\sqrt{\beta m_0^2} \leq 0.2$ (right). 
        A solid curve shows the analytical solution of the Maxwell theory on a lattice in the left panel while the mass perturbation result for $\sqrt{\beta m_0^2} =0.01$ in the right.
        }
    \label{fig:mass_f}
\end{figure}

\begin{figure}[htbp]
    \centering
    \includegraphics[width=0.48\hsize]{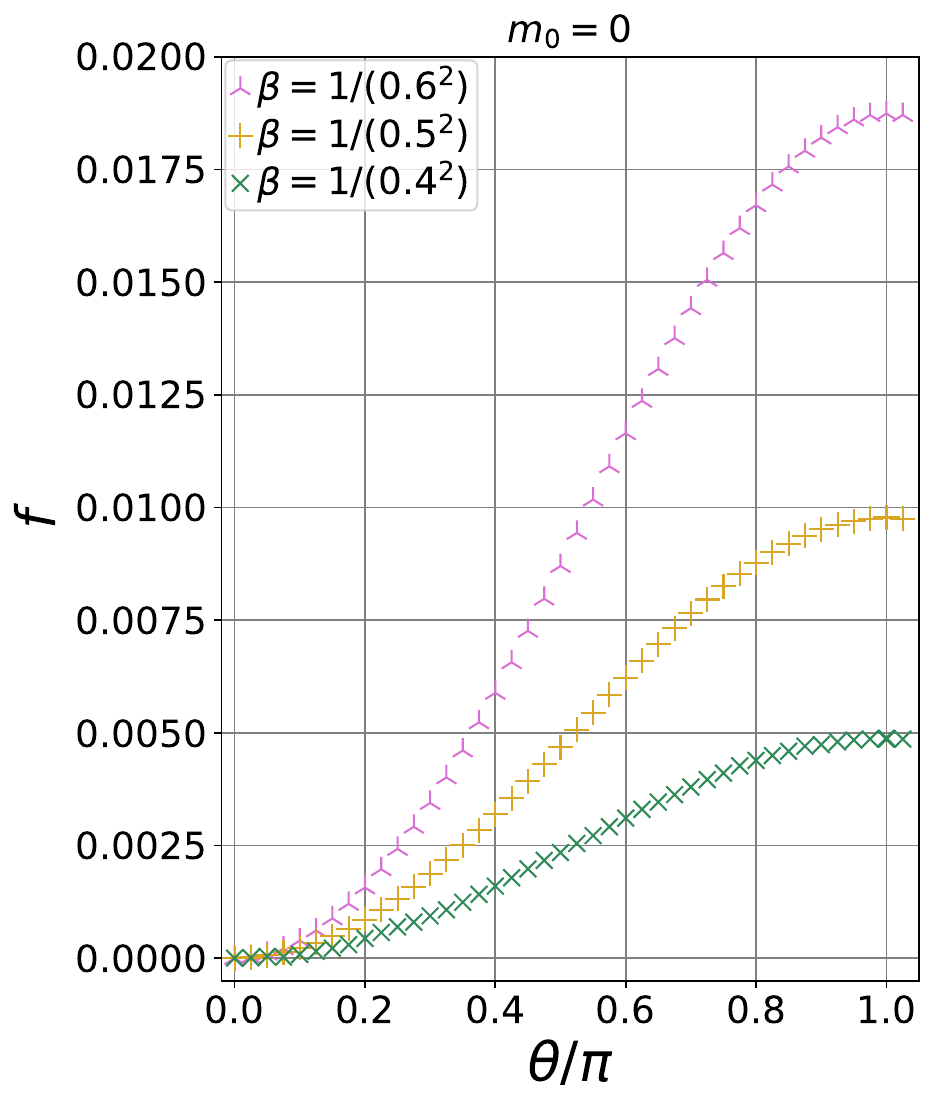}
    \caption{
         Free energy density as a function of $\theta/\pi$ at $\sqrt{\beta m_0^2}=0$ with various $\beta$. Note that the plot of $\beta=1/(0.5^2)=4$ is also depicted in Fig.~\ref{fig:mass_f}.
    }
    \label{fig:f_artifact5}
\end{figure}

\begin{figure}[htbp]
    \centering
    \includegraphics[width=0.48\hsize]{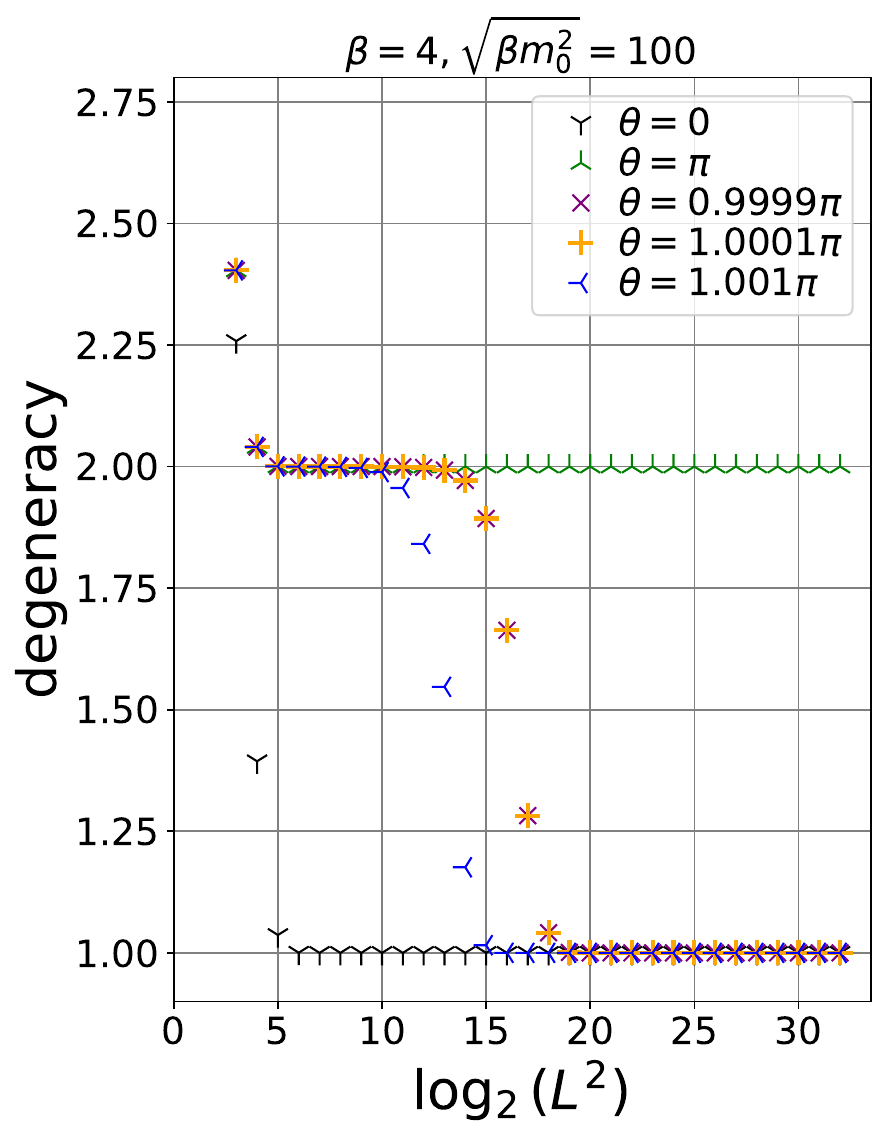}
    \includegraphics[width=0.48\hsize]{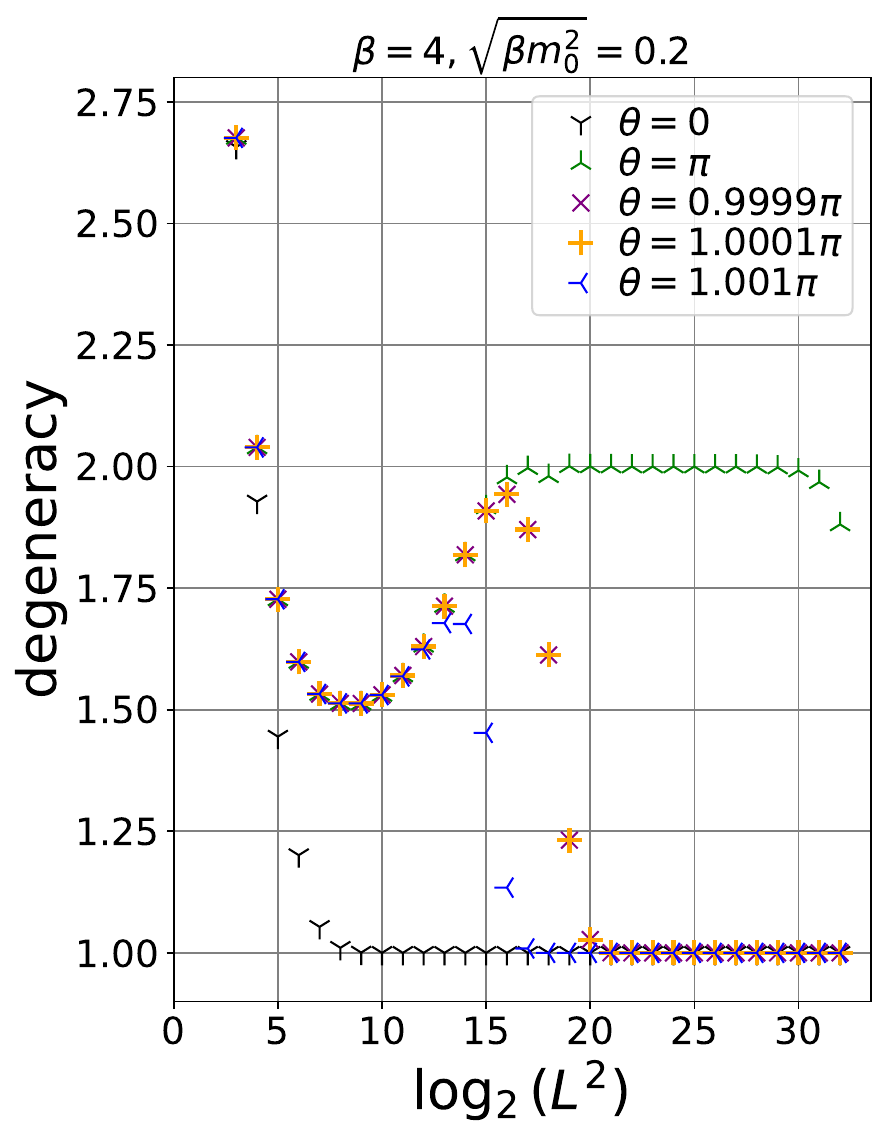}
    \caption{
        Ground state degeneracy as a function of the coarse-graining steps at $\sqrt{\beta m_0^{2}}=100$ (left) and $\sqrt{\beta m_0^{2}}=0.2$ (right).
    }
    \label{fig:degeneracy_bm10000}
\end{figure}

Let us now investigate the finite-mass effect.
Fig.~\ref{fig:mass_f} depicts the free energy density with various masses at $\beta=1/(0.5)^2=4$.
First, our numerical result with $\sqrt{\beta m^{2}_{0}}=100$ is consistent with the large mass limit of the lattice model.
When the mass decreases, the results deviate from the large mass limit, which is nothing but the finite-mass effect.
Also, the free energy density tends to be smooth with respect to $\theta$ at $\theta=\pi$ by further decreasing the mass.
This behavior seems quite similar to that of the $N_{f}=1$ Schwinger model with a $\theta$ term, where the first-order transition terminates at a critical endpoint.
However, we expect that the cusp at $\theta=\pi$ should appear for any finite mass because there is no critical endpoint in the $N_{f}=2$ model in the continuum limit~\cite{Coleman:1976uz, Dempsey:2023gib}.
In addition, although the free energy density should be independent of $\theta$ at $m=0$ in the continuum, the result at $m_0=0$ depends on $\theta$ as shown in the right panel of Fig.~\ref{fig:mass_f}.
Furthermore, the results with $\sqrt{\beta m_0^2}=0.01$ have non-negligible discrepancy from those in the mass perturbation theory, shown as a curve in the right panel of Fig.~\ref{fig:mass_f}. 
These inconsistencies might be due to the finite-$\beta$ effect.
Indeed, the free energy density at $m_0=0$ depends significantly on $\beta$, as shown in Fig.~\ref{fig:f_artifact5}.

We also investigate the degeneracy of the ground state at $\theta=0$ and around $\theta=\pi$ by calculating the so-called fixed-point tensor~\cite{PhysRevB.80.155131}.
Shown in Fig.~\ref{fig:degeneracy_bm10000} are the numerical results of the ground-state degeneracy as a function of the coarse-graining steps in the BTRG.
We observe a clear plateau of $2$ ($1$) at $\theta=\pi$ ($0$) for $\sqrt{\beta m_0^{2}}=100$. 
We also find that the slight change of $\theta$ from $\theta=\pi$ eventually gives the plateau of $1$. 
These behaviors are also seen for $\sqrt{\beta m_0^2}\ge 0.2$.
These indicate that symmetry breaking occurs and that the ground state is doubly degenerate only at $\theta=\pi$ even for a finite large mass.
However, we did not observe ground-state degeneracy for a small mass ($\sqrt{\beta m_0^2}\le 0.14$).
This would be due to the finite-$\beta$ effect in our calculation at $\beta=4$.
Modification of the phase diagram at finite $\beta$ will be examined in future work.

\section{Conclusion}
\label{sec:conclusion}

We investigated the lattice $N_{f}=2$ Schwinger model with a $\theta$ term.
We used the staggered fermion action combined with the $U(1)$ Wilson gauge action in our calculation. 
We adopted the logarithmic form for the $\theta$ term, which guarantees the $2\pi$ periodicity to the $\theta$ parameter.
To formulate the tensor network representation, we employ the Gauss-Legendre quadrature for the link variables and the Grassmann tensor network for the staggered fermions. 
We also used the BTRG algorithm to obtain more accurate numerical results.

Our numerical results of the free energy density show $2\pi$ periodicity for the $\theta$ parameter.
The free energy density was obtained in a broad range of mass.
We confirmed that our numerical calculation at a large mass provided consistent results with those in the large mass limit.
On the other hand, the results at $m_0=0$ have $\theta$ dependence which should be absent in the continuum limit.
We also observed twofold vacua at $\theta=\pi$ and a unique vacuum for the other $\theta$ in the larger mass regime.
However, such a degeneracy was not found in smaller mass regimes.
These suggest that the phase diagram may be changed by the finite $\beta$.
To achieve a more precise examination of the finite-$\beta$ effect, particularly in the smaller mass region, it will be necessary to increase the algorithmic parameters. 
The algorithm such as the randomized SVD~\cite{PhysRevE.97.033310,Morita:2024lwg} will enable us to carry out such studies, which is left for future studies.

\acknowledgments

The numerical results presented in this paper were completed by the computing system of Yukawa Institute for Theoretical Physics, Kyoto University (Sushiki server and Yukawa-21).
HK would like to thank Shigeki Sugimoto and Yuya Tanizaki for their useful discussions.
This work started during the workshop held at Yukawa Institute for Theoretical Physics (YITP-W-22-13).
The work of HK was supported by the establishment of university fellowships towards the creation of science technology innovation and RIKEN Special Postdoctoral Researcher Program.
SA acknowledges the support from JSPS KAKENHI Grant Number JP23K13096, JP24H00214, the Endowed Project for Quantum Software Research and Education, the University of Tokyo (\url{https://qsw.phys.s.u-tokyo.ac.jp/}), the Center of Innovations for Sustainable Quantum AI (JST Grant Number JPMJPF2221), the computational resources of Wisteria/BDEC-01 and Cygnus and Pegasus under the Multidisciplinary Cooperative Research Program of Center for Computational Sciences, University of Tsukuba, and the Top Runners in Strategy of Transborder Advanced Researches (TRiSTAR) program conducted as the Strategic Professional Development Program for Young Researchers by the MEXT.
KM is supported in part by Grants-in-Aid for JSPS Fellows (Nos.\ JP22J14889, JP22KJ1870) and by JSPS KAKENHI with Grant No.\ 22H04917.
ST is supported in part by JSPS KAKENHI Grants No. 21K03531, and No. 22H01222.
This work was supported by MEXT KAKENHI Grant-in-Aid for Transformative Research Areas A “Extreme Universe” No. 22H05251.

\bibliographystyle{JHEP.bst}
\bibliography{bib/kanno,bib/trg_algorithm,bib/trg_formulation,bib/trg_gauge,bib/trg_grassmann,bib/trg_review}


\end{document}